\documentclass[]{eps}

\usepackage{graphicx}
\usepackage{times}

\volumenumber{58}
\publishyear{2006}
\frompage{\pageref{firstpage}}
\topage{\pageref{finalpage}}

\shorttitle{L. Kolokolova \lowercase{\textit{et al.}}: Comet dust as a mixture of aggregates and solid particles}

\title{Comet dust as a mixture of aggregates and solid particles: model consistent with ground-based and space-mission results}
\author{Ludmilla Kolokolova$^1$ and Hiroshi Kimura$^2$}
\affiliation{$^1$Department of Astronomy, University of Maryland, College Park, MD, 20740, USA\\
             $^2$Institute of Low Temperature Science, Hokkaido University, Sapporo 060-0819, Japan\\}
\received{xxxx xx, 2008}\revised{xxxx xx, 2008}\accepted{xxxx xx, 2008}
\abstract{
The most successful model of comet dust presents comet particles as aggregates of submicron grains. It qualitatively explains the spectral and angular change in the comet brightness and polarization and is consistent with the thermal infrared data and composition of the comet dust obtained {\it in situ} for comet 1P/Halley. However, it experiences some difficulties in providing a quantitative fit to the observational data. Here we present a model that considers comet dust as a mixture of aggregates and compact particles. The model is based on the Giotto and Stardust mission findings that both aggregates (made mainly of organics, silicates, and carbon) and solid silicate particles are present in the comet dust. We simulate aggregates as {\bf  Ballistic Cluster-Cluster Aggregates (BCCA)} and compact particles as polydisperse spheroids with some distribution of the aspect ratio. The particles follow a power-law size distribution with the power $-3$ that is close to the one obtained for comet dust {\it in situ}, at studies of the Stardust returned samples, and the results of ground-based observations of comets. The model provides a good fit to the angular polarization curve. It also reproduces the positive spectral gradient of polarization, red color of the dust, and {\bf low albedo. It also has the ratio of compact to fluffy particles close to the one found  {\it in situ} for comet 1P/Halley } and the mass ratio of silicate to carbonaceous materials equal to unity that is in accordance with the elemental abundances of Halley's dust found by Giotto mission. }


\begin{document}
\label{firstpage}
\maketitle
\copyrighttext{}

\section{Introduction}

Comets are bodies that have preserved the pristine materials from the time of the planet formation. Thus, knowledge about the comet dust allows us to find out the properties of the materials in the early Solar System. This is why comets have been objects of special interest not only to ground-based observers but also to numerous space missions: Giotto and Rosetta (European Space Agency), Vega (Soviet Space Agency), Suisei  {\bf (Institute of Space and Astronautical Science, currently a section of Japanese Aerospace Exploration Agency),}
and NASA missions Contour, DS1, Stardust, and Deep Impact. 

Accumulation of the observational data about comet dust (i.e., observations in the comet continuum filters free of gas emissions) reveals how comet dust brightness and polarization change with the phase angle (angle Sun-Comet-Observer) and with the wavelength in the visible and near-infrared spectral range (see details in the review by Kolokolova et al. 2004a). There were numerous attempts to use these data to solve the inverse problem for comet dust, i.e., to find out size, composition, shape, and structure of its particles based on the observational data. Unfortunately, the interpretation usually was done based only on photometric or only polarimetric data or was limited only by the angular dependencies of the observed characteristics. 
{\bf 
In such a limited case, one could find 
}
particles (sometimes rather simple, e.g., spheres or cylinders) whose light-scattering properties fit some observational data. However, any attempts to use the same model of comet dust to interpret other observational data failed (see Kolokolova et al 2004a for review and references). It was shown in Kolokolova et al. (2004b) that as soon as we try to fit photometric and polarimetric angular and spectral data together, no model of compact particles can be applied. Even the most complex model of polydisperse multishaped particles could not fit the whole scope of the observational data no matter what refractive index was tried. Even considering the comet particles as porous by calculating their refractive index using Maxwell Garnett mixing rule and considering voids as part of the mixture (similar to the approach described in Mukai et al., 1992) no ensemble of particles that could reproduce the correct shape of polarization and brightness phase curve together with the correct color and polarimetric spectral gradient (polarimetric color) was found. 

Recently, a significant progress has been achieved in explaining the observational data describing the comet dust as an ensemble of aggregates made of submicron particles (see Kimura et al. 2003, Mann et al. 2004, Kimura et al. 2006). This model allows not only to qualitatively reproduce the observed changes in brightness and polarization with phase angle and wavelength but also can explain existence of two types of comet infrared spectra characterized by strong and weak $10\,\mu$m silicate feature. It was shown {\bf
in Kolokolova et al. (2007) } that the comets characterized by a strong silicate feature in thermal infrared and high polarization have more porous particles, {\bf as it was first pointed out by Li and Greenberg (1998),} whereas the comets characterized by a weak silicate feature and low polarization contain more compact particles. {\bf This model turned out to account for optical and infrared observations of ejecta from comet 9P/Tempel 1 during Deep Impact mission (Yamamoto et al. 2008).} One more advantage of this model is that it is consistent with the data on comet dust composition obtained {\it in situ} for comet  1P/Halley  and with the structure of Interplanetary Dust Particles (IDPs) (Jessberger et al. 2001). However, as explained later, this model does not provide quantitative fits to the observed dependence of polarization on phase angle. Here, we present a more advanced model that account for all the optical properties of comet dust not only qualitatively but also quantitatively.

\section{Ground-based observations and their interpretation}

Despite the difference in size, age, orbit, and source of origin, the vast majority of comets shows very similar photopolarimetric characteristics in the visible {\bf (see Kolokolova et al. 2004a, 2004b and references therein)}. They are the following:
\begin{itemize}
\item{Low geometric albedo of the particles, close to 4--5\%.}
\item{Prominent forward-scattering and gentle back scattering peaks in the angular dependence of intensity, "flat" behavior at medium phase angles} 
\item{For a broad range of wavelengths angular dependence of linear polarization demonstrates  (i) negative branch of  polarization for phase angles  $\le 20^{\circ}$ with the minimum $P_{\rm min} \approx  2$\%;  ii)  bell-shaped positive branch with  low maximum of value $P_{\rm max} \approx 15$--25\%   at  the phase angle in the range 90--$100^{\circ}$.} 
\item{Usually red or neutral color for a broad range of wavelengths that does not change with the  phase angle.}
\item{Polarization at a given phase angle above 30$^\circ$ usually increases with the wavelength (red polarimetric color); this increase  gets larger with the phase angle.} 
\end{itemize}
As it was mentioned in the introduction, all of these observational features could be qualitatively simulated if we consider the comet particles as aggregates of submicron grains. The model works well for both porous (Ballistic Cluster-Cluster Aggregates, BCCA) and more compact (Ballistic Particle-Cluster Aggregates, BPCA) particles. 

Using the T-matrix multisphere code (Mackowski \& Mishchenko 1996) Kimura et al. (2006) accomplished a survey of light-scattering properties of aggregates. The best fit to the comet observational data was achieved for the case of the aggregates of $\approx 0.2$ micron-size particles made of a mixture of organics, carbon, silicates with a small admixture of {\bf iron sulfide or }
iron. {\bf The last component was added based on detection of FeNi metals in IDPs of likely cometary origin 
 and carbonaceous meteorites (Bradley et al., 1996; Bernatowicz et al. 1999).
 The presence of FeNi metal in cometary dust is
confirmed by the results of the Stardust mission (Zolensky et al. 2006). Such a model could also reproduce a low albedo equal to 4--5\% that is close to the one found from the numerous observations and space mission studies. However, a quantitative fit to the observed degree of polarization 
seemed to require very big, made of hundreds of thousand monomers, aggregates. In other case the computational results showed too small values of the negative polarization and too high degree of the maximum polarization. Such huge aggregates require enormous computer time and memory to calculate their light-scattering characteristics. Also, domination of large  fluffy aggregates in comet dust may be unrealistic assumption  at large distances %especially far 
from the nucleus where fragile aggregates get smaller due to fragmentation. Thus, the model definitely requires some improvements which we attempt to provide in this paper based on additional information obtained from the space-mission data.}

\section{Comet dust as viewed by space missions}

The model of aggregates described in the previous section was not developed in isolation from the results of the space mission to comets. First, the aggregate model of comet dust partly was stimulated by studies of IDPs captured in the upper layers of the Earth's atmosphere. These studies showed that IDPs of presumably cometary origin are aggregates of submicron particles (Brownlee et al. 1980). Second, Kimura et al. (2003) used the composition of comet dust consistent with the average composition of the dust in comet 1P/Halley measured {\it in situ} by Giotto mission (Jessberger et al. 1988). Hereafter we will call this composition Halley-like. In the next section we describe it in more detail.

Recent space missions have added new information to the known properties of the comet dust. The most significant insight was provided by the Stardust mission. This is not surprising since this was the mission specifically designed to study the comet dust {\it in situ} and to bring back samples of comet dust. The {\it in-situ} measurements by instruments DFMI and CIDA allowed revealing the size distribution of the particles (Green et al. 2004; see in more detail below) and provided some information on the composition of the dust (Kissel et al. 2004). However, the most important findings have been done during the study of the returned samples of the comet dust.

The returned samples present the comet dust particles captured when a special panel made of aerogel cells was exposed to the dust flux in the coma. Study of the tracks of the dust particles in the aerogel and in the aluminum foil around the cells revealed the structure of the particles. The majority of them appeared to be fragile aggregates of small {\bf submicron particles}. Their composition was complex and included both silicate and carbonaceous materials as this was found for comet 1P/Halley. However, besides the aggregates, rather large non-agglomerated, solid silicate particles were also found among the Stardust returned samples (see, e.g., Flynn 2008).  {\it In-situ} study of comet 1P/Halley showed that dust in comet 1P/Halley consisted of three types of particles (Fomenkova et al. 1999):  rock (silicate), CHON (organic) and mixed (averaged, Halley-like dust composition) particles. The rock particles were shown to be solid whereas mixed and organic were fluffy (Jessberger \& Kissel, 1991).  Both rock and mixed particles were found in the Stardust returned samples.{\bf The absence of pure organic particles in Stardust samples is not surprising. It is very likely that the high-speed interaction of the particles with the aerogel caused evaporation of small organic particles. Notice that organic-rich particles were detected at in-situ measurements by Stardust CIDA instrument (Kissel et al. 2004). }

In the next section we consider a model of comet dust that includes the space-mission results discussed in this section. Besides the fact that the model considers the comet dust as a mixture of solid silicate and aggregated organic and Halley-like particles, the particle size distribution should be also consistent with the one found by cometary space missions. 

There have been several sets of space-mission data that allowed determining size distribution of comet particles. H\"{o}rz et al. (2006) summarizes the comet dust size distribution obtained from the Stardust studies of tracks in aerogel and craters in the aluminum foil, Stardust DFMI {\it in-situ} measurements of comet 81P/Wild 2 dust, and Giotto DIDSY and PIA measurements of dust in comet 1P/Halley. All size distributions look very much alike and in all cases consist of three power-law size distributions: least steep for the particles smaller than 1 micron, steeper for the particles of size between 1 and 100 micron and even steeper for the larger particles. The exponent of the power law for all the cases is negative and changes within 2--4 that is close to the values obtained from ground based observations, e.g., observations of comets C/1996 C1 Tabur 
{\bf and 174P/Exhclus (Kolokolova et al. 2001; Bauer et al. 2008). }

\section{Combined model of comet dust}

The combined model of comet dust outlined in the previous section is combined in two meanings. First, it combines the results of successful interpretation of ground-based observations with the {\it in-situ} space mission data and laboratory studies of the returned samples of comet dust. Second, it combines particles of two structural types (aggregates and solid particles) and three types of composition (silicate particles, organic particles, and particles of Halley-like composition).  

We consider light-scattering by the solid silicate particles using the T-matrix method for spheroids by Mishchenko et al. (1996). {\bf The refractive index equal to $1.6+0.001i$ that we use in the calculations is close to the values of optical constants for glass of olivine or pyroxene composition
reported by Dorshchner et al. (1995).} The particles are presented by multishaped, polydisperse mixture of spheroids. Size distribution of the particles is selected to be consistent with the results of studies of comet dust by the space missions described at the end of the previous section. The power of the size distribution of the dust particles has been found negative in a range of the values from 2 to 4 and the exact value depends not only on the size of particles but also on the method used to study the size distribution and on the position of the instruments in the coma. This is why we decided not to follow the found size distributions in detail but consider a single power law with the power equal to $-3$, which represents 
the average value among the found ones. Multishaped properties of the ensemble were simulated by considering a mixture of oblate and prolate spheroids with the axis ratio within the range 1--2.5.
 
Aggregates in the combined model were represented as Ballistic-Cluster-Cluster Aggregates (BCCA) of organic and Halley-like composition. Aggregates were made of 256 monomers of radius 0.1 $\mu$m. The light-scattering properties of these aggregates were calculated using the multisphere T-matrix code by Mackowski \& Mishchenko (1996). In case of organic aggregates we use the refractive indices from Li \& Greenberg (1997). The Halley-like composition, as it was mentioned above, represents average composition found at {\it in-situ} measurements of the dust in comet 1P/Halley (Jessberger et al. 1988). It represents a mixture of silicates, metals, and carbonaceous materials. We suppose that one third of the carbonaceous materials are in the form of organic refractory and two thirds are in the form of amorphous carbon. This is based on a comparison of the nitrogen abundance between Halley's and interstellar dust (Kimura et al 2003). {\bf We consider that silicates, iron, and organics refractory are embedded in amorphous carbon and estimate the average refractive index of the mixture using the multi-component Maxwell Garnett mixing rule (Bohren \& Huffman 1983, Chapter 8.5).} Taking into account the elemental abundances of Halley's dust the volume filling factors are 31.76\% for silicates, 2.56\% for iron, and 65.68\% for carbonaceous materials. The refractive indices of iron, organics refractory, and amorphous carbon are taken from Johnson \& Christy (1974), Li \& Greenberg (1997), and Rouleau \& Martin (1991), respectively. {\bf For silicates we adapt the refractive index of astronomical silicate from Laor \& Draine (1993). This material has slightly higher absorption than such silicates as olivine and pyroxene that allows us to simulate a slight admixture of iron sulfide found in the comet dust (Zolensky et al. 2006, Flynn 2008).} 

{\bf We computed the intensity in the scattered plane $I_{par}$ and in the perpendicular plane $I_{per}$ separately for polydisperse multishaped spheroids, for organic BCCA, and for Halley-like BCCA, and then calculated the final intensity $I=I_{per}+I_{par}$ and polarization $P=(I_{per}-I_{par})/I$ summarizing $I_{per}$ and $I_{par}$ for all constituents of the mixture with some factors that were varied to control the silicate-to-carbonaceous-materials ratio and to adjust our results to the observational data.}

The best fit results are shown in Fig.1 where the top picture shows the change of albedo at two wavelengths, 0.45 and 0.6 $\mu$m, and the bottom one shows the phase-angle change in polarization. One can see that we have managed to fit the cometary photopolarimetric data rather well. {\bf The brightness follows the phase angular dependence described in Section 2, albedo at zero phase angle is small, about 3\% for the red filter, and the brightness at the shorter wavelength is smaller than at the longer one, i.e., the color is red. } The calculated polarization also fits the angular dependence known from the observations of comets, and not only qualitatively but also quantitatively. {\bf Even though we get a deeper negative polarization branch that was observed, with the minimum $\le 4$\%, it changes the sign at approximately 25$^{\circ}$, and reaches the maximum values at about 25--30\% at the phase angle around 95$^\circ$ that fits very well to the observational data.} Also, the polarization at the shorter wavelength is smaller than at the longer one (red polarimetric color) and the difference between them increase with the phase angle that correctly reproduces the observational trends. 

As we have mentioned above, we have one more parameter that was controlled during our calculations. This is the mass ratio of silicate to carbonaceous materials. The result presented at Fig.1 corresponds to this ratio equal to unity that is consistent with the elemental abundance of the comet Halley dust (Jessberger et al. 1988). Finally, the best-fit result appears to be in accordance with the abundance of rock, organic, and mixed particles found at {\it in-situ} measurements of the dust in comet 1P/Halley.  We found that the best fit is achieved if the ratio of rock particles to the organic ones and to the Halley-like aggregates is equal to 0.30 : 0.44 : 0.26 that is consistent with the results by Fomenkova (1999) who showed that the rock particles represented 1/3 of the Halley dust. Our model also agrees with the fact that the rock particles in the Halley dust were compact whereas other particles were fluffy (Jessberger \& Kissel 1991).

{\bf The presented combined model shows a significant improvement in fitting the whole scope of the ground-based and space-mission data for comet dust. The further development of the model (e.g., by including larger size of particles and larger axes ratio for compact grains) is planned to get a better quantitative fit to the polarization and albedo values.}

%% one-column wide figure
%%
\begin{figure}[Fig1]
\centerline{\includegraphics[width=12.5cm,clip]{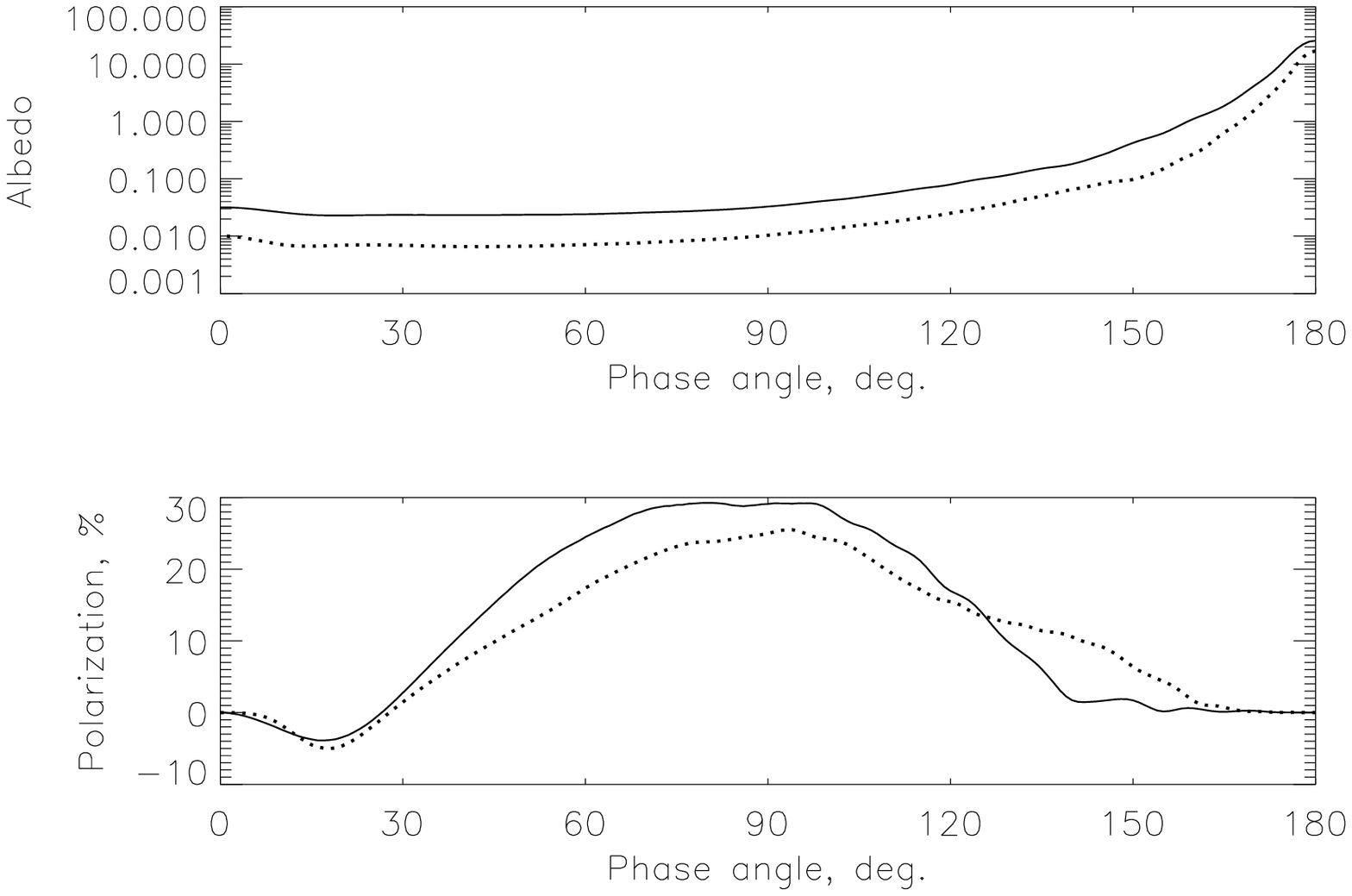}} 
%%\centerline{\includegraphics[scale=.,clip]{file name}}
\caption{Dependence of albedo (top) and polarization (bottom) on the phase angle for the mixture of the aggregates of organic and Halley-like composition and compact silicate particles described in Section 4.  Solid line is for the wavelength 0.6 $\mu$m and dotted line is for the wavelength 0.45 $\mu$m}

\end{figure}

\section{Conclusions}

The model of comet dust that combines two types of particles: aggregates of submicron particles of organic and Halley-like composition and compact silicate particles can correctly reproduce all observational data and fits to the data on albedo,  photometric phase curve, polarimetric phase curve, color, polarimetric color and ratio of silicates and organics in the comet dust. It confirms the results obtained {\it in situ} for the dust in comet 1P/Halley and is consistent with recent Stardust findings. This model could provide a much better fit to the photometric and polarimetric observational data than the model that considered only aggregates.  The model correctly reproduces the polarimetric data, {\bf including a pronounced negative polarization at small phase angles } and the positive polarization with the maximum value less than 30\% at the phase angle around  $95^{\circ}$ and red polarimetric color. It also shows red color of comet dust, albedo of the dust {\bf equal to $\approx 3$\%,} and the mass ratio of silicate to carbonaceous materials in the dust equal to unity that is in accordance with the elemental abundances of Halley's dust. Finally, it shows the ratio of rock, organic, and mixed particles in the dust that agrees with the Giotto {\it in-situ} measurements.

\acknowledgments{
\bf 
This work is in part supported by the Japanese Ministry of Education, Culture, Sports, Science, and Technology {\sl MEXT} (Monbu Kagaku Sho) under Grant-in-Aid for Scientific Research on Priority Areas.}

%%\lastpagecontrol{20cm}

\email{L. Kolokolova (e-mail: ludmilla@astro.umd.edu)}
\label{finalpage}
\lastpagesettings
\end{document}